\documentstyle[12pt]{article}
\newcommand{\CaCa}{Ca~$+$~Ca~}

\newcommand{\vr}{\mbox{\boldmath $r$}}

\newcommand{\vp}{\mbox{\boldmath $p$}}

\title{\bf  
Relativistic Effects in the Transverse Flow 
in the Molecular Dynamics Framework}

\author{ Tomoyuki Maruyama$^{1}$, Koji Niita$^{1,2}$, Toshiki Maruyama$^{1}$,
\\ Satoshi Chiba$^{1}$, Yasuaki Nakahara$^{1}$ and Akira Iwamoto$^{1}$ \\
\\[1ex]
$^1$ Advanced Science Research Center, \\
Japan Atomic Energy Research Institute,\\
Tokai, Ibaraki 319-11, Japan \\[1ex]
$^2$ Research Organization for Information Science and Technology, \\
Tokai, Ibaraki 319-11, Japan }

\date{}

\begin{document}

\maketitle

\begin{abstract}
In order to investigate relativistic effects we compare  the transverse flow
calculated by using the four versions of the QMD approaches with 
that of the full covariant RQMD approach.
From the comparison we conclude that the simplified RQMD (RQMD/S), 
which uses the common time coordinate to all particles, 
can be used instead of  RQMD up to 6 GeV/u.
\end{abstract}

\vfil
\eject
\newpage

The Quantum Molecular Dynamics (QMD) approach is one of the most powerful 
models to describe heavy-ion \cite{Aich} and light-ion \cite{Niita} reactions 
in the several tens to the several hundreds MeV/u energy region.
In the high energy region above about 1 GeV/u this approach is also useful
though relativistic effects become significant.
The relativistic effects mean not only the relativistic kinematics,
but also the Lorentz covariance of the interactions.

In the relativistic energy region we have the following problems 
when we introduce the relativistic kinematics and the Lorentz transformation 
into the non-covariant QMD approach. 
First, the increase of the initial density due to the Lorentz contraction 
makes an additional repulsion through the density-dependent force, 
which causes the spurious excitation and 
the unphysical instability of initial nuclei.
Second, we cannot correctly evaluate the internal energies of fast-moving 
fragments at the end of the QMD calculation 
because the non-relativistic mean-field used in QMD 
is variant under the Lorentz transformation.
In fact, these effects clearly appear in the transverse flow \cite{MARU2,Leh}
and in the multiplicity of alpha particles in the heavy-ion collisions 
even at $E_{\rm lab} \sim$ 1 GeV/u \cite{TOMO2}.
At the relativistic energy, therefore, the Lorentz covariant transport 
approach is desirable to make all nuclei and fragments hold 
the consistent phase-space distribution under the Lorentz transformation.

The Relativistic QMD (RQMD) approach \cite{Sorge,MARU1} is the most
useful theoretical model for this purpose; it is formulated to describe 
the interacting $N$-body system in a fully Lorentz covariant way based on
the Poincare-Invariant Constrained Hamiltonian Dynamics \cite{PICHD}.
The position and momentum coordinates of the $i$-th particle, $q_i$ and $p_i$, 
are defined as four-dimensional dynamical variables and the functions of
the time evolution parameter $\tau$.
The on-mass-shell constraints are given by
\begin{equation}
H_i \equiv p_i^2 - m_i^2 - 2 m_i U_i = 0 ,
\end{equation}
where $m_i$ and ${U_i(q_j,p_j)}$ are the mass and the Lorentz scalar 
quasi potential for the $i$-th particle. 
The explicit form of the quasi-potential is determined by 
the requirement that it corresponds to the non-relativistic 
mean-field in the low energy limit \cite{Sorge,MARU1}.
Whereas in the non-relativistic framework the argument of the potential 
is a square of the relative distance $\vr^2_{ij}$ between two particles, 
in RQMD we take it as a square of the relative distance 
in the rest frame of their CM system as
\begin{equation}
- q_{{\rm T}ij}^2 = 
- q_{ij}^2 + \frac{ ( q_{ij} \cdot p_{ij} )^2 }{ p_{ij}^2 }
\label{arg1}
\end{equation}
with
\begin{eqnarray}
q_{ij} =  q_i - q_j , & p_{ij}  =  p_i + p_j .
\label{arg2}
\end{eqnarray}

In this formulation the time coordinate $q_i^0$ is distinguished from 
the time evolution parameter $\tau$ and constrained by the following
time-fixation as
\begin{equation}
\chi_i \equiv
\sum_{j (\neq i)} 
\frac{e^{q_{{\rm T}ij}^2/L_c}}{q_{{\rm T}ij}^2/L_c} q_{ij} p_{ij}
=0 .
\label{tfix}
\end{equation}
Under this time-fixation the colliding two particles have equal values of
their time coordinate at the their CM frame in the dilute gas limit.

Using the above on-mass-shell and the time-fixation constraints
\cite{Sorge,MARU1}, we can describe the Lorentz covariant motions of 
particles.
These constraints are chosen to be completely consistent 
in the non-relativistic framework at the non-relativistic limit
of $m_i \rightarrow \infty$.

The above expression includes some important features of the Lorentz
covariance as follows.
First, the quasi-potential is given as the Lorentz scalar while 
it is given as a time coordinate of the Lorentz vector in the
usual QMD approach.
Second, the change from $\vr_{ij}^2$ to $- q_{{\rm T}ij}^2$
causes the direction dependent forces.
The attractive force in the fast moving nuclei is stronger in the
moving direction than in the transverse direction.
The above two effects keep the Lorentz contracted 
phase-space distribution of fast moving matter stable.

The mean-field affects only  low energy  behaviors in the reaction.
Without the Lorentz covariance in the mean-field, however, we cannot 
correctly give intrinsic motions of nucleons in the fast moving nuclei.
Namely the above expression can describe relatively low energy
phenomena in the fast moving system.

There are, however,  two kinds of problems in applying the RQMD approach to 
the analysis of experimental data.
One is the fact that it needs too long CPU time, and
the other is the energy conservation problem.
In RQMD, it is not easy to satisfy the energy conservation
after the meson production and absorption.
The change of the particle number due to the meson production and absorption
breaks the time fixation.
Thus we have to resolve the values of $q_i^0$ and $p_i^0$ to impose 
the constraints (the on-mass shell condition and the time-fixation) again.
This procedure leads to the discontinuity of particle coordinates and 
causes the change of the potential energy.
Particularly the latter problem is very serious to study the fragmentation 
process because IMF multiplicities are dominantly determined by the excitation 
energy of the residual nucleus.

In order to avoid these difficulties we propose the simplified version 
of RQMD (RQMD/S) in the following way.
Equation (\ref{tfix}) gives the time-fixation requiring 
equal time coordinates of two colliding particles in their center-of-mass 
system in the dilute gas limit.
However a choice of the time fixation is completely arbitrary except for
the non-relativistic limit and the cluster separability \cite{Sorge}. 
We thus adopt the time fixations to equalize the time evolution parameter 
$\tau$ and all time coordinates of baryons and mesons as
\begin{equation}
\chi_{i}  \equiv  {\hat a} \cdot ( q_i - q_N )  = 0 ~~~~ ( i = 1 \sim N-1 ) ,
\end{equation}
where ${\hat a}$ is a four-dimensional unit-vector taken 
as $(1;\vec{0})$ in the reference frame.
By this simplified choice of the time fixations, 
the energy conservation after the change of particle number is always satisfied.
Though the time-coordinate has a physical meaning only in the reference frame,
the above time-fixations are still defined in particle-motions 
Lorentz covariantly by the mean-field.

Of course these time-fixations break the Lorentz covariance in the two-body
collisions at the ultra-relativistic energy \cite{Sorge}.
In our work, however, we are interested in the phenomena around several GeV/u,
where our simplification does not make any significant troubles \cite{GyWolf}.

To reduce the computation time, furthermore, we make an approximation
that the momentum coordinate $p^0_i$ in the argument of the quasipotential is 
replaced by the kinetic energy 
${\varepsilon}_i = \sqrt{ {\vp}_i^2 + m_i^2 }$.
This approximation does not affect the final results much because 
the quasi-potential is much smaller than the kinetic energy 
in the relativistic energy region. 

Now we must examine the above simplified  RQMD (RQMD/S) before applying it to
the data analysis.
For this purpose we arrange the five kinds of methods as follows.
The first is the standard QMD, so called QMD/G, 
which does not include the Lorentz contraction of the initial distribution.
The second is the QMD/L, which is QMD including the Lorentz contracted
initial distribution.
The third is the QMD/R, where the mean-field is treated as a time-component of
the vector type, but the argument is varied along the eqs. (\ref{arg1},
\ref{arg2}).
The forth is the RQMD/S which is the simplified RQMD explained above. 
The last one is the full RQMD.
Then we calculate  the directed transverse momentum, which is 
most sensitive to the relativistic effects \cite{MARU2}, defined as
\begin{equation}
<{\bf P}^{{\rm dir}}_x> = {1\over A} \sum_{i=1}^A {\rm sign} \left[ 
              Y_{{\rm CM}}(i) \right] p_{x}(i),
\end{equation}
where $Y_{CM}(i)$ and $p_{x}(i)$ are the center-of-mass rapidity and the
transverse momentum in the reaction plane of the $i-$th nucleon. 

In the numerical calculations
the predictor-corrector method is used to integrate the equation of motion. 
For the two-body effective interaction, 
we use a Skyrme-type interaction with HARD EOS 
(the incompressibility $K$ = 380 MeV). 
The widths of the wave packets are taken from the values for \CaCa 
in Ref.~\cite{Toshi2}.
The initial nuclei are given with the cooling method \cite{Cooling}.
We omit the Coulomb force and two-body collision term 
for a simplicity because our purpose is the examination of the
relativistic effects.

In Fig.\ \ref{fig-flow} we show the energy dependence of 
$<{\bf P}^{{\rm dir}}_x>$,.
We plot the result of the QMD/G, QMD/L, QMD/R and RQMD/S simulation 
as a difference from that of RQMD \cite{MARU1}
for $^{40}{\rm Ca}$ + $^{40}{\rm Ca}$ reactions
at $b$ = 2~fm, for the energy range from 150 MeV/u 
to 6 GeV/u.

As mentioned before, the Lorentz contraction of the initial 
phase space distribution increases the flow, which is shown by 
the change from the dot-dashed line to the dashed line. 
By the full covariant treatment of the interaction, however, 
this effect is counterbalanced with the Lorentz covariance of
the mean-field, but still remains an increased flow 
from QMD/G value \cite{Leh}. 

As seen in Fig.\ \ref{fig-flow}, 
the prescription of QMD/R does not deviate so much from the full covariant 
treatment up to 3 GeV/u. 
At much higher energy, however, the result of QMD/R is decreasing linearly 
from that of RQMD. 
On the other hand, RQMD/S gives results very similar to RQMD up to
6 GeV/u and more.

The difference between the RQMD (RQMD/S) and QMD/R comes from the 
different treatment of the potential; 
a Lorentz scalar type in the former, while a time-component of the 
vector type in the latter, respectively. 
This is understood qualitatively by considering 
a single particle motion under a fixed external potential $U$. 
In the Lorentz scalar treatment of the potential $U$, 
the single particle energy $p_i^0$ is expressed in this simple 
case as
\begin{equation}
p_i^0 = \sqrt{\vp_i^2 + m_i^2 + 2m_i U}.
\end{equation}
Accordingly the equation of motion is 
\begin{equation}
\dot{{\vp}}_i = - \sum_{j}
     \frac{m_j}{p^0_j} \frac{\partial U_j}{\partial {\vr}_i}.
\end{equation}
On the other hand, in QMD/R they are 
\begin{equation}
p_i^0 = \sqrt{\vp_i^2 + m_i^2 } + U_i,
\end{equation}
and 
\begin{equation}
\dot{{\vp}}_i = - \sum_{j} \frac{\partial U_j}{\partial {\vr}_i}.
\end{equation}
The form of ${\partial U_j}/{\partial \vr_i}$, 
which is attractive at the early time stage of nucleus-nucleus collisions, 
is almost the same in the QMD/R and in the RQMD (RQMD/S).
In the high energy region the single particle energy $p^0_i$ has 
approximately a same value for all nucleons in the early time stage.
Thus the force of QMD/R becomes larger and deviates linearly 
from that of the RQMD (RQMD/S) as energy increases. 
Above 3 GeV/u, hence, the difference between the  Lorentz scalar
and a time component of the Lorentz vector becomes significant,
and the full covariant prescription for the mean-field is necessary 
to describe the reactions, particularly the nucleus-nucleus collisions. 

In summary we have calculated the directed transverse momentum 
using the five kinds of (R)QMD approaches and have
discussed their relativistic effects.
It has been shown that the Lorentz covariance of the mean-field is very 
important above about 1 GeV energy region.
The RQMD/S can give almost the same results as the full RQMD in the transverse
flow which is thought to be the most sensitive observables 
to the relativistic effects at present.
Hence we can conclude that the RQMD/S approach can be used up to 6 GeV 
instead of the full RQMD approach.



%
%
\begin{figure}
\caption{The directed transverse momentum as a function of energy per nucleon 
for $^{40}{\rm Ca}+^{40}{\rm Ca}$ reaction at $b$ = 2~fm.
The results are shown as the differences from that of the RQMD 
\protect\cite{MARU1}.
The dot-dashed line with open circles, the dashed line with open boxes, 
the dotted line with full boxes and the solid line with full boxes 
denote the results of QMD/G, QMD/L, QMD/R and RQMD/S, respectively. 
        }
\label{fig-flow}
\end{figure}

\begin{thebibliography}{10}

\bibitem{Aich} J. Aichelin, Phys. Rep. {\bf 202} (1991) 233, and
reference therein.
\bibitem{Niita} K, Niita, S. Chiba, T. Maruyama, T. Maruyama, H. Takada,
T. Fukahori, Y. Nakahara and A. Iwamoto, Phys. Rev. {\bf C52} (1995)
2620.
\bibitem{MARU2} T. Maruyama, G.Q. Li and A. F\"assler,
Phys. Lett. {\bf 268B} (1991) 160.
\bibitem{Leh} E. Lehmann, R.K. Puri, A. F\"assler, T. Maruyama,
G.Q .Li, N. Ohtsuka, S.W. Huang, D.T.Khoa, Y.Lotfy, M. A. Matin,
Prog. Part. Nucl. Phys. 30 (1992) 219
\bibitem{TOMO2} T. Maruyama, T. Maruyama and K. Niita,
Phys. Lett. {\bf B358} (1995) 34.
\bibitem{Sorge} H. Sorge, H. St\"ocker and W. Greiner, Ann. of
Phys. {\bf 192} (1989) 266.
\bibitem{MARU1} T. Maruyama, S.W. Huang, N. Ohtsuka, G.Q. Li, A. F\"assler 
and J. Aichelin, Nucl. Phys. {\bf A534} (1991) 720.
\bibitem{PICHD} For Example, P. A. M. Dirac, Rev. Mod. Phys. 
{\bf 21} (1949) 392; \\
A. Komer, Phys. Rev. {\bf D18} (1978) 1881, 1887, 3617 ;\\
J. Samuel, Phys. Rev. {\bf D26} (1982) 3475, 3482. 
\bibitem{GyWolf} Gy. Wolf, G. Batko, W. Cassing, U. Mosel,
K. Niita and M. Sh\"affer, Nucl. Phys. {\bf A517} (1990) 615
\bibitem{Toshi2} T. Maruyama, A. Ohnishi and H. Horiuchi,
Phys. Rev. {\bf C45} (1992) 2335.  
\bibitem{Cooling} T. Maruyama, A. Ohnishi and H. Horiuchi, Phys. Rev
{\bf C42} (1990) 386. 

\end{thebibliography}
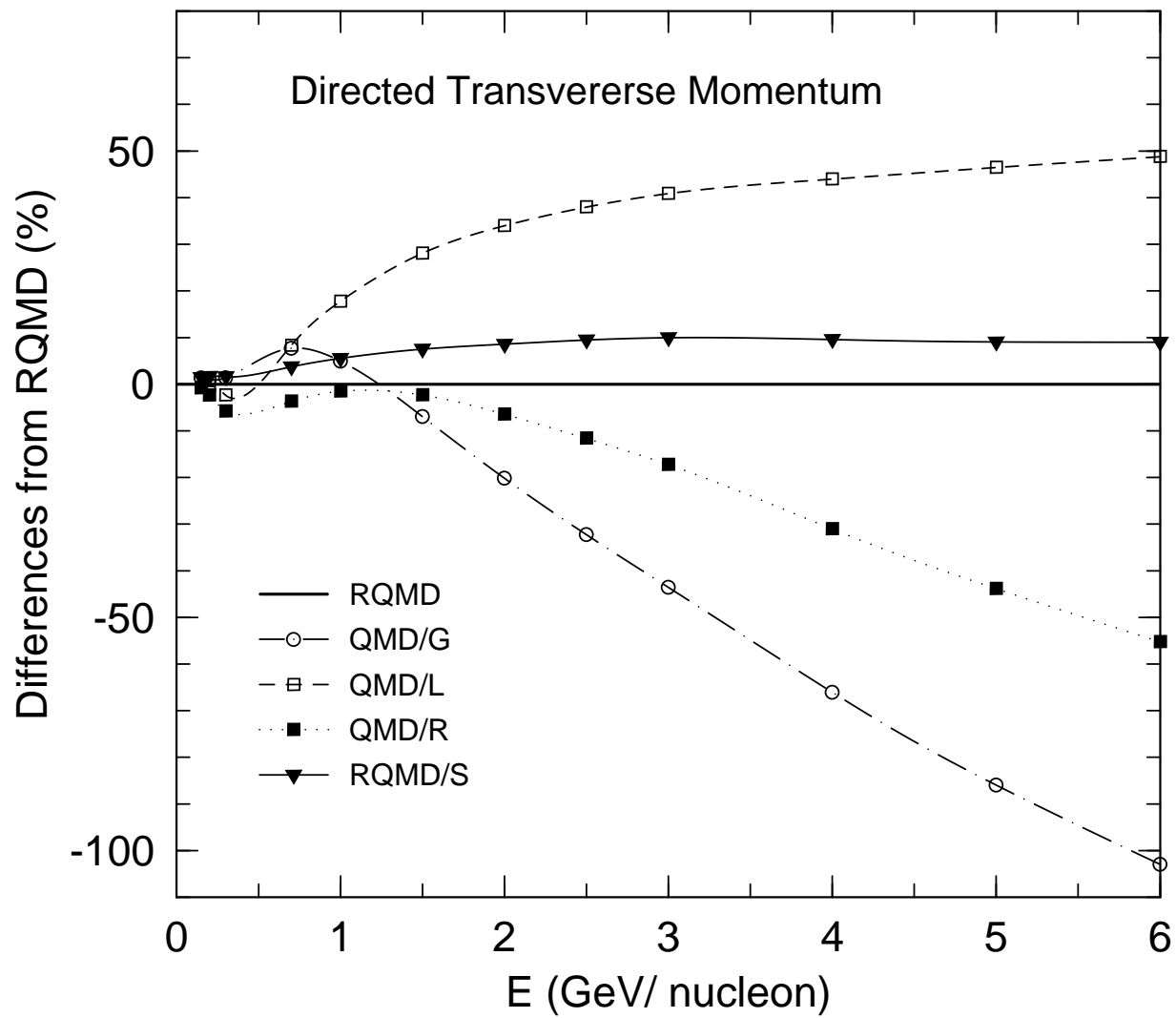
\end{document}